\documentclass[aps,prl,twocolumn,showpacs,superscriptaddress]{revtex4}
\usepackage{amsmath} 
\usepackage{amssymb} 
\usepackage{epsfig}
\newcommand{\p}{\partial} 
\newcommand{\dd}{\hbox{d}}
 
\newcommand{\ii}{\hbox{i}}

\newcommand{\half}{\text{$\frac{1}{2}$}}

\newcommand{\bpsi}{\bar{\psi}} 
\newcommand{\bphi}{\bar{\phi}}

\newcommand{\Tr}{\text{Tr}}
\newcommand{\begeq}{\begin{equation}}
\newcommand{\eqend}{\end{equation}}
\begin{document}

\title{Non Perturbative Renormalization Group study of reaction-diffusion processes}
\author{L\'eonie Canet}
\affiliation{Laboratoire de Physique Th\'eorique et Hautes Energies, Universit\'{e}s Paris VI Pierre et Marie Curie, Paris VII Denis Diderot, 2 place Jussieu, 75251 Paris cedex 05, France}
\author{Bertrand Delamotte}
\affiliation{Laboratoire de Physique Th\'eorique et Hautes Energies, Universit\'{e}s Paris VI Pierre et Marie Curie, Paris VII Denis Diderot, 2 place Jussieu, 75251 Paris cedex 05, France}
\author{Olivier Deloubri\`ere}
\affiliation{Department of Physics, Virginia Polytechnic Institute and State University, Blacksburg, Virginia 24061-0435, USA}
\author{Nicolas Wschebor}
\affiliation{Instituto de Fisica, Facultad de Ingenieria, J.H.y Reissig 565, 11000 Montevideo, Uruguay}

\begin{abstract}
We generalize  nonperturbative renormalization group methods to nonequilibrium critical phenomena. Within this formalism,  reaction-diffusion processes
are described by a scale dependent effective action, which flow is derived.
We investigate  branching and annihilating random walks with odd number
of offsprings. Along with recovering their universal physics (described by the directed percolation universality class), we determine their phase diagrams and predict that a transition  occurs even in three dimensions, contrarily to what perturbation theory suggests.  

\end{abstract}

\maketitle
\date{\today}

The description of many physical, chemical, and biological systems resorts to nonequilibrium many-particle models.
Their dynamical evolution is governed by a Markov process ruled by microscopic transition probabilities which violate 
detailed balance. The ensueing macroscopic behaviors are hence intrinsically much richer than in equilibrium.
Many models exhibit phase transitions between stationary states, which have unveiled the existence of universal properties.
 The characterisation of the nonequilibrium  universality classes constitutes a prevalent issue
of statistical mechanics. In this respect, reaction-diffusion processes  provide a simple and thus valuable framework to
investigate nonequilibrium critical phenomena. They generically describe diffusing particles that can undergo different 
microscopic reactions  \cite{hinrichsen00,tauber02}.
  The most common universality class encountered is that of directed percolation 
(DP) \cite{grassberger78,janssen81}. It can be simply 
achieved through a reaction-diffusion model endowed with the processes:
\begin{eqnarray}
&2A&\to\varnothing \hspace{0.5cm} \hbox{pair annihilation at rate $\lambda$}
\label{death}\\
&A&\to 2A  \hspace{0.35cm} \hbox{generation of an offspring at rate $\sigma$}
\label{birth}\\
&A&\to\varnothing \hspace{0.5cm} \hbox{spontaneous decay at rate $\mu$} 
\label{decay}
\end{eqnarray}
 The competition between birth and death gives rise to 
a steady state where the density saturates at a constant value $\rho$. The system 
exhibits a continuous phase transition between an active (fluctuating) state where $\rho > 0$,
 and an inactive (absorbing) state where the density  vanishes.
 In fact, a large variety of  models, like the contact process \cite{harris74}, Reggeon Field Theory
 in particle physics \cite{moshe78},
 the dimer poisoning \cite{ziff86},
the  autocatalytic reaction models \cite{aukrust90}, show the same critical behavior.

Despite considerable successes, the general understanding of nonequilibrium phenomena suffers from the lack of a  
 systematic theoretical framework. Indeed, out of equilibrium, the absence of a free-energy functional 
from which to draw the probability distribution of steady states prevents from the use of  well-established equilibrium methods. 
Instead, the master equation should be solved, which is not feasible in general.
 Fortunately, it can often be cast into a path integral representation
 \cite{doi76}, rooting a field theory that can  be 
probed with a dynamical renormalization group (RG).
 The usual perturbative techniques consist in expansions in series
of a parameter. They have allowed to achieve numerous
advances in our theoretical understanding. Nevertheless, as they rely on the smallness of 
a parameter, their validity remains confined to the vicinity of critical dimensions or to weak coupling regimes.

 We propose in this Letter a powerful alternative, based on a nonperturbative
 RG procedure (NPRG) \cite{tetradis94,ellwanger94c,morris94a}. 
This method avoids the previous  pitfalls as it does not intrinsically depend on the magnitude of a parameter.
Moreover, the microscopic interactions can be traced all along the NPRG flow, and related to the long-range properties,
 thus giving access to
 non universal quantities, such as  equations of state and  critical temperatures \cite{berges02}. 
In this work, we  generalize the NPRG methods to nonequilibrium systems and derive very generic, model-independent RG flow equations for reaction-diffusion processes. We exploit these equations to investigate branching and annihilating random walks (BARW) with odd number of offsprings. We first recover their universal physics which belong to the DP universality class. Then, linking the density $\rho$ of
stationary states with the microscopic rates  
enables us to  push further  a recent analysis by Cardy and T{\"a}uber  \cite{cardy96}. We predict in particular that a phase transition can occur in three dimensions, contrarily to what was conjectured. 

The NPRG formalism --- the effective average action
method \cite{tetradis94} --- is a continuous implementation, on successive momentum shells, 
of Wilson's block-spin concept.
For equilibrium systems, it consists in building a sequence of effective actions $\Gamma_k$, at the  
 running  momentum scale $k$, interpolating between the  microscopic Hamiltonian $H$ of the system 
and its Gibbs free energy
$\Gamma$. This is achieved by adding to the partition function ${\cal Z}$ a mass-like term 
 $\Delta H_k$ that suppresses the  
 propagation of the low-energy modes (with wave number $q<k$) while not affecting
the high-energy ones ($q>k$). The corresponding free energy log${\cal Z}_k$,
 or its Legendre transform $\Gamma_k$, thus only encompasses the large momentum fluctuations. 
At the scale $k=\Lambda$,
$\Lambda^{-1}$ corresponding to the underlying  lattice spacing, no  fluctuation has  yet been  taken into account
and $\Gamma_{k=\Lambda} = H$, while at $k=0$, all  fluctuations have been integrated out and $\Gamma_{k=0} = \Gamma$.
An exact RG equation for the flow of $\Gamma_k$ is easily derived and underlies the NPRG formalism \cite{tetradis94}.
We now generalize this formalism to out of equilibrium systems, for which
partition function and Gibbs free energy no longer exist. Despite this lack,
 a path integral formulation of the dynamics can be set up, commonly involving an auxiliary response field $\bphi$, 
 along with the usual field 
$\phi$. It allows to define a generating functional ${\cal Z}$ for the correlation functions, and one, $\Gamma$, for the one-particle 
irreducible correlation functions. These functionals constitute the building-blocks of an  out of equilibrium NPRG.
We introduce in ${\cal Z}$ a scale-dependent term, analogous to $\Delta H_k$:
\begin{equation}
{\cal Z}_k[\bar{J},J]=\int D\phi D\bphi\ e^{ -S[\bphi,\phi]-\Delta S_k[\bphi,\phi] +  \int J\,\phi + \int \bar{J}\,\bphi},
\end{equation}
where
\begin{equation}
\Delta S_k = \int \frac{\dd^d p}{(2\pi)^d} \frac{\dd \omega}{2\pi}\, \Phi(-p,-\omega)\, \hat{R}_k(p)\,\Phi^T(p,\omega),
\end{equation}
 $\Phi$ denoting the two-component vector  $(\bphi,\phi)$.
The $2\times 2$ matrix $\hat{R}_k$ is  symmetric  with zeros on its diagonal,
 and a mass-like function  $R_k(p)$ off its diagonal. 
 The selection of modes  is ensured by imposing that, on the one hand, $R_k(p)\sim k^2$
 for $p\ll k$, that is $R_k$ acts as an effective mass --- an IR cut-off --- for slow modes. 
On the other hand, one imposes that $R_k(p)\to 0$ for $p\gg k$, so that the rapid modes remain unaltered. In this work,
we use $R_k(p)= (k^2-p^2)\theta(k^2-p^2)$  \cite{litim01} which gives simple analytical expressions for the momentum integrals.  
 We emphasize that
no cutoff is necessary for time, since the corresponding integrals are regular and the frequency dependence 
can be integrated over (see below). The effective action 
$\Gamma_k$ is defined through the appropriate Legendre transform
 $\Gamma_k$ + log${\cal Z}_k =  \int J\psi+ \bar{J}\bpsi - \Delta S_k(\bpsi,\psi)$,
where $\psi$  (resp. $\bpsi$) is the mean value of $\phi$ (resp. $\bphi$), defined as the derivative of
 ${\cal Z}_k$ w.r.t. $J$ (resp. $\bar{J}$). The mass term $\Delta S_k$ 
 ensures that the proper limits  $\Gamma_{k=\Lambda} = S$ and  $\Gamma_{k=0} = \Gamma$ are recovered. 
 The exact flow equation of $\Gamma_k$ is then easily derived, similarly to the equilibrium calculation \cite{tetradis94}: 
\begin{equation}
\p_k \Gamma_k (\bpsi,\psi)=\half \Tr \left\{ [\hat{\Gamma}_k^{(2)}+\hat{R}_k]^{-1}\p_k \hat{R}_k\right\},
\label{flow}
\end{equation}
where $\hat{\Gamma}_k^{(2)}$ is the field dependent matrix of the second 
functional derivatives of $\Gamma$, and Tr stands for matrix trace over internal indices and integration
 over the internal momentum and frequency. 
Eq. (\ref{flow}) is   a  functional 
 equation  which obviously cannot  be  solved exactly.   To
handle it, one has to {\it truncate} $\Gamma_k$. 
Since the critical physics corresponds to the long distance ($q\to 0$), long time ($\omega \to 0$) limit,
a sensible truncation  consists  in
expanding  $\Gamma_k$ in powers of  gradients \cite{tetradis94} and time derivatives.
Retaining only  the leading order in derivatives,  the functional $\Gamma_k$ for reaction-diffusion systems reads:
\begin{equation}
\Gamma_k(\bpsi,\psi)=\int\dd^d r \, \dd t\, [ Z_k \bpsi (\p_t - D_k \nabla^2)\psi\\
 + V_k(\bpsi,\psi)].
\label{GammaDP}
\end{equation}
In principle, the three functions $V_k$, $Z_k$ and $D_k$ depend on the fields, and their functional form 
 is dictated by the symmetries of the underlying dynamics. We study in the following the leading order approximation
that consists in neglecting  the field-dependence 
of  $Z_k$ and $D_k$~\cite{tetradis94}. Thus, we only consider their  $k$-dependence, which 
 accounts for the critical anomalous scalings of the fields, and of time.
Indeed, the anomalous dimension $\eta$ of the fields is defined such that at criticality,
 $\psi$ and $\bpsi$ scale as $k^{(d+\eta)/2}$. 
 This definition induces that $\eta=-\p \ln Z_k /\p \ln k$. The exponent ($z-2$) embodies the anomalous scaling between time
 and space in the critical scaling regime \cite{wijland98}. It follows that  $2-z=-\p \ln D_k /\p \ln k$.
We emphasize that the previous {\it ansatz} (\ref{GammaDP}) 
should qualitatively well encompass the physics of the model, and moreover 
provide accurate estimates of ``static'' quantities, related to the potential part.  
However, this level of approximation is not expected to give very accurate dynamical exponents, even less that they are large.
 Refining these exponents would require to 
include the field dependence of $Z_k$ and $D_k$, and as a further step, to enrich the derivative content
 of (\ref{GammaDP}) \cite{berges02,morris98d,canet03a}. 

We can now establish the generic flow equations for reaction-diffusion processes. 
The flow of $V_k$ is drawn from  Eq.\,(\ref{flow}), evaluated at a uniform and stationary field configuration.
After integrating over $\omega$, one gets:
\begin{equation}
\p_k V_k = \frac{1}{2Z_k} \int\frac{\dd^d p}{(2\pi)^d}\,\frac{\p_k R_k}{\sqrt{1-\displaystyle{\frac{\p^2_{\psi}V_k\p^2_{\bpsi}V_k}{h_k(p)^2}}}},
\label{Flow}
\end{equation}
where $h_k(p)= Z_k D_kp^2 + R_k(p) + \p^2_{\bpsi\psi}V_k$.
The flow of  $Z_k$  (resp. $D_k$) is obtained taking the $q^2$ (resp. $\nu$) part of the flow of $\p^2_{\bpsi,\psi}\Gamma_k(p,\nu)$.
These flows depend on the fields, but the resulting exponents should be identical for any field values. 
However, this property 
breaks down when any truncation is performed \cite{berges02}.
 Since, in the reaction-diffusion models, $\bphi$ vanishes,
 and the average of $\phi$ reaches a  uniform and stationary configuration, it is sensible to
 define the exponents at $\bpsi=0$ and $\psi(p,\nu)=\psi_{0k}\delta^d(p)\delta(\nu)$. 
 Derivating twice Eq.\,(\ref{flow}) w.r.t. $\bpsi(-q,-\nu)$ and $\psi(q,\nu)$ yields,  after integrating over the internal frequency:
\begin{widetext}
\begin{eqnarray}
\p_k \p^2_{\bpsi\psi}\Gamma_k(q,\nu)& =& \frac{1}{Z_k}\p^3_{\bpsi\psi\psi}V_k\p^3_{\bpsi\bpsi\psi}V_k\int\frac{\dd^d p}{(2\pi)^d}\, \frac{\p_k R_k(p)}{[Z_k\ii\nu +h_k(p)+h_k(p-q)]^2}\nonumber   \\
& &-\frac{1}{2 Z_k}\p^2_{\bpsi\bpsi}V_k(\p^3_{\bpsi\psi\psi}V_k)^2\int\frac{\dd^d p}{(2\pi)^d}\,\p_k R_k(p)\left( \frac{h_k(p)+ h_k(p-q)}{h_k(p)h_k(p-q)[Z_k \ii \nu + h_k(p)+h_k(p-q)]^2}\right.\nonumber \\
& &\left. + \frac{1}{h_k(p)^2[Z_k \ii \nu + h_k(p)+h_k(p-q)]}\right).
\label{FlowGam}
\end{eqnarray}
\end{widetext}


As a first step, we show that these equations allow to recover the universal physics of DP
and more precisely to compute the three usually quoted independent critical exponents 
 $z$, $\nu$ and $\beta$. $\nu$ describes the scaling behavior of the
correlation length near criticality, $\beta$ that of the order parameter.
 A field theoretical 
action  $S[\bphi,\phi]$ can be derived from the microscopic
 processes (\ref{death}), (\ref{birth}) and (\ref{decay}), upon the introduction of two fields, $\phi$ and  $\bphi$.   
After their rescaling and a shift of $\bphi$, one gets  \cite{cardy96}: 
\begin{multline}
S[\bphi,\phi]=\int \dd^d r \, \dd t
\, \left[\bphi(\p_t - D \nabla^2)\phi + (\mu - \sigma)\phi\bphi  \right.\\ \left.+
\sqrt{2\lambda\sigma}(\bphi\phi^2-\phi\bphi^2) + \lambda\phi^2\bphi^2\right],
 \label{dp}
\end{multline}
which is invariant under the simultaneous changes
 $\phi(x,t) \to -\bphi(x,-t)$ and $\bphi(x,t) \to -\phi(x,-t)$.
The generic term of the effective potential $V_k(\bpsi,\psi)$ allowed by this symmetry reads
 $a_{\alpha\beta}(\psi^\alpha\bpsi^\beta + (-1)^{\alpha+\beta}\bpsi^\alpha\psi^\beta)$,
which can be expressed in terms of the two invariants $x = \psi\bpsi$ and $y =\psi-\bpsi$.
Thus $V_k$  can be conveniently parametrized by $U_k(x,y) = V_k(\bpsi,\psi)$. By  numerically integrating the flows (\ref{Flow}) and (\ref{FlowGam}), we find that for a fine tuned initial 
($\mu - \sigma$), corresponding to criticality,  the (dimensionless) effective potential  flows to a fixed function.
 The exponents are calculated in the vicinity of this fixed solution.
We first implement the well-known local potential approximation (LPA), that is  $Z_k = D_k =1$.
The static exponent $\nu$ is very accurate in all dimensions,
 see Table \ref{Table}, whereas $\beta$, deduced from the scaling relation $\beta=1/2(d+\eta)$,
 is more imprecise since $\eta=0$ at this order.
We then study the leading order (LO) in derivatives: $Z_k$ and $D_k$ now run with $k$.
 The critical exponents are all the more accurate that  $\eta$ and ($z-2$) remain small. They 
thus deteriorate as the dimension lowers, since  $\eta$ and ($z-2$) then become large. 
As usual with this method, more accurate determinations of $\eta$ and $z$ rely on a proper description of the momentum 
and frequency dependence of the two-point correlation function, which 
would require to implement a richer {\it ansatz} for $\Gamma_k$ \cite{canet03a}, as already emphasized.
\begin{table}
\begin{tabular}{|c|c|c|c|c|}
\hline 
\ \ d\ \ \  &\quad \quad &\ \ (a) LPA \ &\ \ (b) LO \ \  &\ \ (c) MC \ \ \\ \hline 
    &$\nu$    & 0.584 & 0.548 & 0.581(5)\\
3   &$\beta$  & 0.872 & 0.782 & 0.81(1)\\
    &$z$      & 2     & 1.909 & 1.90(1)\\ \hline
    &$\nu$    & 0.730 & 0.623 & 0.734(4)\\
2   &$\beta$  & 0.730 & 0.597 & 0.584(4)\\
    &$z$      & 2     & 1.884 & 1.76(3)\\ \hline
    &$\nu$    & 1.056 & 0.888 & 1.096854(4)\\
1   &$\beta$  & 0.528 & 0.505 & 0.276486(8)\\
    &$z$      & 2     & 1.899 & 1.580745(10)\\ \hline
\end{tabular}
\caption{Critical exponents of DP. (a) and (b): NPRG calculations from this work, at the local potential approximation, resp. at leading order.
(c): from Monte Carlo simulations \cite{jensen99}.}
\label{Table}
\end{table}

We finally come to the most important part of this work: the study of non universal features of the BARW  
 defined by the set of the two processes (\ref{death}) and (\ref{birth}) ($\mu=0$).
Mean field theory predicts the density to reach a finite saturation value for any positive birth rate $\sigma$, 
and thus the system to be always in the active state. Nevertheless, simulations  \cite{tretyakov92}
have brought out the existence of a phase transition in this model in $d=1$ and $d=2$, invalidating mean 
field conclusions. 
 Cardy and T{\"a}uber have addressed this question through a field theoretical perturbative study  \cite{cardy96} and provided a partial answer relevant for  small reaction rates $\lambda/D$ and $\sigma/D$, and in the vicinity of $d=2$.
They  have
shown that  fluctuations can indeed induce, for a non zero critical birth rate $\sigma_c$, a 
dynamical transition to an absorbing state (belonging to the DP universality class).  
The underlying reason for the existence of the transition is that  the combination 
of  (\ref{death}) and (\ref{birth})  generates under renormalization the one-particle spontaneous decay (\ref{decay}). 
This process allows the existence of an absorbing state if its renormalized rate  $\mu_R(\sigma,\lambda)$
 renders $\mu_R - \sigma_R$ positive (see Eq.~\ref{dp}). In $d=2$, this defines a narrow absorbing region 
below a critical transition curve given by:
\begin{equation}
\sigma_c(\lambda)/D \propto e^{-\displaystyle{4\pi D/\lambda}}.
\label{sig}
\end{equation} 
For $d>2$, perturbation theory breaks down. However, since the transition curve Eq.~(\ref{sig}) is already infinitely flat for small $\lambda/D$ in $d=2$, and since fluctuations are expected to be suppressed when the dimension increases, it has been suggested in \cite{cardy96} that no transition should exist in $d=3$.

\begin{figure}[ht]
\includegraphics[width=55mm,height=80mm,angle=-90]{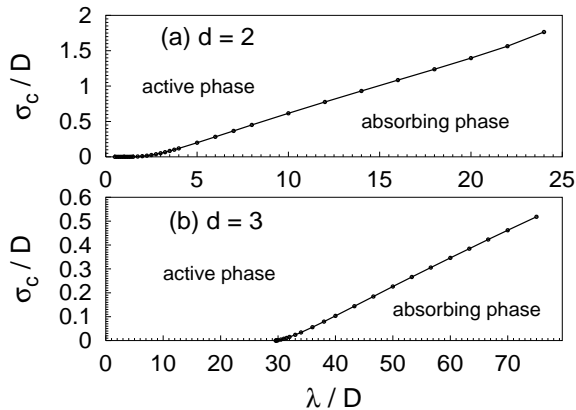}
\caption{Phase diagrams of BARW in $d=2$ and $d=3$.}
\label{perc}
\end{figure}

We re-investigate this question using NPRG, which is neither  restricted to small $\lambda/D$ nor to a 
specific dimension. We can thus  explore the phase diagrams of BARW in $d=2$ and $d=3$. 
They  are displayed in Fig.~(\ref{perc}).

%

In $d=2$, we find two regimes. For small $\lambda/D$ (typically below $4$) the transition curve 
is very accurately fitted by an  exponential with a coefficient $-11.8$, in agreement with Eq.~(\ref{sig}) at a $6\%$ level. 
For large $\lambda/D$, $\sigma_c(\lambda)/D$ grows linearly  opening up a large absorbing phase.
 In $d=3$, we also find two regimes, one with a linear critical curve, and one without transition, separated by a threshold value 
 $\lambda/D\simeq 30$. Let us now give simple  arguments highlighting  why a transition must exist in $d=3$, that perturbation theory misses.

When  $D$ is large compared with $\lambda$ and $\sigma$, the mechanism of particle destruction is driven by the  pair annihilation occuring when 
 wandering particles meet. While in $d=2$ all the particles can be destroyed through this mechanism, it is not  efficient enough  to do so in $d=3$.
 We thus expect only an  active phase  at large $D$ in $d=3$. This corresponds to the result of  the perturbation theory, which is indeed valid at small $\lambda/D$.
On the other hand, in the small $D$ regime where perturbation theory fails, the mechanism  of destruction is dominated by ``self-destruction'',
 that is particles annihilate with their own offsprings before they can diffuse. This process no longer requires the random encounter of particles
 and thus does not depend on the dimension. We therefore expect a transition to occur at small $D$ both in two and  three dimensions.
 These arguments corroborate the phase diagrams obtained from the NPRG analysis. 
Though, relying on the {\it ansatz} (\ref{GammaDP}), we cannot provide a quantitative  estimate of the accuracy of the critical rates in Fig. \ref{perc}, the previous arguments reinforce the  NPRG ability  to capture the non universal physics of the model over the different diffusive regimes. Moreover, we have performed numerical simulations, which confirm the existence of a DP phase transition in $d=3$ for large $\lambda/D$, and hence support our prediction (extensive simulations to quantitatively check the phase diagrams are in progress~\cite{canet04a}).

 Note that in simulations \cite{tretyakov92}, the implementation of reaction (\ref{birth}) involves the creation of offsprings on neighboring sites, contrarily to the on-site birth process underlying field theoretical approaches (both \cite{cardy96} and ours). Though this should not alter the behavior of the model when $D$ is large, it becomes crucial at small $D$ since the ``self-destruction'' mechanism is then drastically suppressed. This is probably why no transition is found in $d=3$ in these simulations.

In the present work, we have implemented for the first time NPRG methods in
 nonequilibrium statistical physics, and derived  very generic flow equations for reaction-diffusion models.
 This method  has allowed us to recover the universal behavior of 
 directed percolation and moreover to compute the phase diagrams of BARW. We have shown  that
 a phase transition exists in $d=3$ in these systems contrarily to what suggests perturbation theory. 
The formalism developed here offers an as efficient as promising mean of investigation of nonequilibrium physics.
 This could give some valuable insight into the perennial Kardar-Parisi-Zhang equation,
describing  roughening transitions in growing interfaces.
\begin{acknowledgments}
We wish to thank  H. Chat\'e for providing us codes and assistance for simulations and F. Van Wijland and U. C. T\"auber for helpful discussions. The LPTHE is Unit\'e Mixte du CNRS, UMR 7589.
\end{acknowledgments}


\end{document}